# Fast Electrical Control of Single Electron Spins in Quantum Dots with Vanishing Influence from Nuclear Spins

J. Yoneda, [1,2] T. Otsuka, [1,2] T. Nakajima, [1,2] T. Takakura, [1] T. Obata, [1] M. Pioro-Ladrière, [3] H. Lu, [4†] C. Palmstrøm, [4] A. C. Gossard, [4] and S. Tarucha [1,2*]

[1]Department of Applied Physics, University of Tokyo, 7-3-1 Hongo, Bunkyo-ku, Tokyo 113-8656, Japan,

[2]RIKEN, Center for Emergent Matter Science, Hirosawa 2-1, Wako-shi, Saitama 351-0198, Japan,

[3]Département de Physique, Université de Sherbrooke, Sherbrooke, Québec J1K 2R1, Canada,

[4]Materials Department, University of California, Santa Barbara, California 93106, USA.

* Corresponding author. tarucha@ap.t.u-tokyo.ac.jp

† Currently at College of Engineering and Applied Sciences, Nanjing University, Nanjing, Jiangsu 210093, China

**Abstract**

We demonstrate fast universal electrical spin manipulation with inhomogeneous magnetic fields. With fast Rabi frequency up to 127 MHz, we leave the conventional regime of strong nuclear-spin influence and observe a spin-flip fidelity > 96%, a distinct chevron Rabi pattern in the spectral-time domain, and spin resonance linewidth limited by the Rabi frequency, not by the dephasing rate. In addition, we establish fast *z*-rotations up to 54 MHz by directly controlling the spin phase. Our findings will significantly facilitate tomography and error correction with electron spins in quantum dots.

In quantum spintronics [1-3], the ability to electrically control electron-spins with high accuracy plays an essential role. Such control in nanoscale devices is widely performed by electron-spin-resonance (ESR), also commonly utilized to investigate the magnetic environment in solids [4,5]. One prominent platform for spintronic devices [1-3] is quantum dots (QDs), which are promising candidates for implementing quantum bits (qubits) [6-9] due to their long coherence time [10-12] and potential for scalability. Indeed, recent experiments based on GaAs QDs have demonstrated two elementary building blocks for universal quantum operations: encoding spin-1/2 qubits by ESR [13-15], and manipulating the two-spin entanglement [16,17]. However, slow control of single spins poses limitations on scaling quantum circuits.

Decoherence is a common enemy of spintronics and quantum computation. For solid-state electron spins, the predominant interaction with the environment to induce decoherence is the hyperfine coupling [18,19]. As the number of nuclear spins is numerous (~ $10^6$ in GaAs QDs [7]), its effective field is approximately Gaussian-distributed with a standard deviation $\sigma$ (= 5 to 10 MHz in GaAs QDs [11,12,20]). Despite rapid advances in this field, ESR rotation commonly act on timescales comparable to the phase coherence time, $T_2^* = 1/(\sqrt{2}\pi\sigma)$, ~50 nsec in GaAs QDs [11] and ~10 nsec in InAs [22], InSb [23] and carbon-nanotube QDs [24]. Therefore, the driven electron-spin dynamics suffers significantly from nuclear spins, which invalidates the Markovian-Bloch equations [25,26]. This hinders precise, coherent spin control. To realize spin-based quantum computation in QDs, it is crucial to reach the *fast* regime, where the single-qubit Rabi frequency $f_{\text{Rabi}} \gg \sigma$, since rapid, sub-nanosecond two-qubit operations are already established [16]. This would be important also in the ESR spectroscopy to reveal coherent spin dynamics, since the problem of hyperfine-induced decoherence is common in materials with abundant magnetic nuclei.

In this Letter, by utilizing distributed magnetic fields, we raise $f_{\text{Rabi}}$ to $\gg \sigma$ (the *fast* regime), and virtually decouple ESR and coherent electron spin manipulation from the nuclear-spin bath. We experimentally reveal the generic relation between the ESR spectrum and $f_{\text{Rabi}}$ in the *slow* to *fast* Rabi regime, and observe a clear difference of driven spin dynamics between the two regimes. In addition, we establish an electrical technique to directly control the spin phase, and achieve phase rotations on similarly short timescales. In contrast to two-axis control through ESR [11,22,23], rotations around the *z*-axis have not been realized before.

To realize the electric control knobs for single-spin rotations, we utilize two types of local magnetic fields induced by a micro-magnet (MM) under an external magnetic field $B_{\text{ext}}$ along the *z*-axis. The first kind is a field gradient $b_{\text{sl}}$ ($= \Delta B_x/\Delta z$), which enables ESR rotation for an electron oscillating in the QD [27]. $f_{\text{Rabi}}$ would be proportional to the product of $b_{\text{sl}}$ and the amplitude of a microwave (MW) that oscillates the electron. The second is the inhomogeneous Zeeman field parallel to $B_{\text{ext}}$. We will show that under such a field, a phase shift can be induced when one swiftly displaces an electron in the QD using pulsed electric fields. Our MM (Fig. 1(a)) is tailored such that both in- and out-of-plane components of the stray field are heavily slanted and their gradients depend only moderately on any geometrical misalignment between the QD and MM (typically of 50 to 100 nm) [28,29]. In order to enhance the effect of the MM, we employ a shallow 2DEG (57 nm below the surface) at an n-AlGaAs/GaAs heterointerface. In the numerical simulation we see that $b_{\text{sl}} \sim 1.2$ mT/nm while $\Delta B_z/\Delta y \sim 0.3$ mT/nm, and $\Delta B_z/\Delta z \sim 1.0$ mT/nm (Fig. 1(b), (c)).

We first characterize the local Zeeman effect due to $B_z$ by measuring ESR for two electrons in the double QD (DQD) as a function of $B_{ext}$ and MW frequency $f_{MW}$. ESR occurs when $f_{MW}$ is equivalent to the Zeeman energy. When we tune the DQD at the (1,1) charge state in Pauli spin blockade (PSB) [30], the blockade is lifted by ESR to generate a transition of the (2,0) charge state. We detect this change in the charge by measuring the change $\Delta G_{QPC}$ of the trans-conductance through a nearby quantum point contact. Two $\Delta G_{QPC}$ peaks due to ESR for the two dots are clearly observed at two different $B_{ext}$ separated by 80 mT or 440 MHz (Fig. 2(a)). This separation is more than 5 times larger than in the previous MM-ESR experiments [15,17], indicating enhancement of the field inhomogeneity ($\Delta B_z/\Delta y$) by the same factor. Assuming a typical value of 100 nm for the DQD inter-dot distance, $\Delta B_z/\Delta y \sim 0.8$ mT/nm, a value consistent with the simulation. Note the large peak separation allows the two spins to be addressed independently even for ESR exceeding 120 MHz.

To further characterize the inhomogeneous field we measure two ESR peaks at various gate voltage points as a function of $B_{ext}$ at a fixed $f_{MW}=8.2$ GHz (Fig. 2(b)-(d)). After the PSB initialization is established, a MW burst is applied at various bias points A to D and 1 to 4 in the (1,1) Coulomb blockade in a pump-and-probe (PP) manner. The two ESR peak positions shift with gate voltage while remaining separated by 80-100 mT, reflecting local $B_Z$ changes. Referring to Fig. 1(c), we estimate that the electron position shifts by ~ 10 nm from A to D or from 1 to 4.

Now we focus on one of the two peaks and examine the Rabi oscillation using a PP technique. After initializing the two spins in the parallel configuration via PSB, we apply a MW burst for $t_{MW}$. Coherent oscillations are observed in $\Delta G_{QPC}$ or finding probability of a flipped spin (Fig. 3(a)). $f_{Rabi}$ increases linearly with MW amplitude up to

~100 MHz and then progressively saturates to ~130 MHz (Fig. 3(b)). One possible scenario that explains this behavior is the anharmonicity of the in-plane QD confinement [31]. We can estimate the fidelity for the initial π-flip to be 96.6 % for the 123 MHz Rabi oscillation with the spin-flip time $t_\pi$ = 4.1 nsec [32]. Although the spin-orbit interaction can drive ESR [33,14], we speculate that our Rabi oscillations are mainly driven by the MM field, since the oscillation behavior cannot be well explained in terms of the relatively weak spin-orbit interaction in GaAs. In the case of very strong MW, we observe faster decay of the Rabi oscillation, which might be explained by photon assisted tunneling (PAT) processes [34,35]. A large MW field enhances leakage to non-qubit states by absorption of photons, which increases for longer burst times, resulting in faster decay of the Rabi oscillation.

Next we examine unique features of ESR in the *fast* Rabi regime. Figure 3(c) shows the ESR line-width vs. $f_{Rabi}$ obtained in the present experiment. Since in the *slow* Rabi limit spins can flip only on exact resonance, the conventional ESR line shape is simply governed by the Gaussian nuclear fluctuation, and this is utilized to extract $T_2^*$ or $\sigma$. For instance, a few mT ESR line-width is reported previously in GaAs QDs with $f_{Rabi}$ below a few MHz [17]. In contrast, in the *fast* Rabi regime the Lorentzian profile is expected with the full-width-at-half-maximum (FWHM) of approximately $2f_{Rabi}$. In this regime with large $f_{Rabi}$, small fluctuations along the *z*-axis hardly affect the ESR rotation axis, as it is determined by the vector sum of contributions. In Fig. 3(c), the FWHM starts gradually increasing for $f_{Rabi}$ > 10 MHz and grows almost linearly with $f_{Rabi}$ for $f_{Rabi}$ > 20 MHz. The line shape and FWHM at the transition from the Gaussian to the Lorentzian profile should be described by those of the Voigt profile. From least-square fitting, we obtain the Landé g-factor, |g| = 0.29 and $T_2^*$ = 61 ns ($\sigma$ = 3.7 MHz), which is

consistent with previous measurements [11,16]. We note that this analysis could underestimate $T_2^*$, since $B_{ext}$ sweep can pump DNP [36,37]. DNP seems to be pronounced under continuous-wave (CW) excitation, although $B_{ext}$ is always swept downwards.

In the *slow* Rabi regime, influence of nuclear spins appears pronounced in Rabi oscillation profiles. It is known that when driven spin dynamics are influenced by a nuclear-spin bath and become non-Markovian, Rabi oscillations are shifted in phase by ~ $\pi/4$ and damped following a power-law [25,26]. This is featured by a rapid damping for the initial spin-flip peak followed by a relatively slow damping. Indeed this feature holds for all previous ESR work in semiconductor QDs [13-15,17,21-25]. On the other hand, the *fast* Rabi oscillation shows large initial oscillations with no $\pi/4$ shift followed by a Gaussian damping of the oscillation. We find that when $f_{Rabi} \gtrsim 40$ MHz the oscillations are well fit by the *fast* Rabi expression for at least up to $6\pi$ spin flips (Fig. 3(d)), whereas the oscillation with $f_{Rabi} \lesssim 15$ MHz is well approximated by the *slow* Rabi expression (Fig. 3(e)).

To further confirm the difference between the *fast* and *slow* Rabi oscillations, we measure the spin dynamics under ESR driving in the time-spectral domain. In Fig. 3(f), a "chevron" interference pattern of the fast Rabi oscillation intensity is clearly recognized as a function of $t_{MW}$ and $B_{ext}$. This is a direct proof that driven spin states remain isolated from the magnetic environment on the timescale of interest. Otherwise, the chevron patterns are smeared due to ensemble averaging over sizable Overhauser fluctuations within the integration time for each data pixel, and scattered due to slow drift within the whole measurement time (Fig. 3(g)). The drift of the center of the interference pattern is also caused by DNP which is comparable to the fluctuation. We discover that both these effects are minimal or absent for the *fast* Rabi oscillation.

Although ESR is sufficient for arbitrary single-qubit control, most quantum circuits contain phase gates, rotations around *z*. Conventionally, a 3-step sequence is assumed for a phase shift by an angle *α*, Z(*α*) = Y(-π/2)X(*α*)Y(π/2). However, the implementation would be much simpler and faster, if this can be directly achieved by electrical gating (the phase of the qubit spin is defined with respect to the reference frame rotating at the frequency of the ESR driving field to allow for ESR π rotations, so it is static under a fixed magnetic field). We show that this is indeed possible by utilizing an inhomogeneous magnetic field. Recalling that the Larmor precession rate is proportional to the local magnetic field of $B_{ext}+B_z$, it depends on the electron position in the dot. Figure 2(d) indicates that $B_z$ can be electrically modulated over the range of 12 mT, corresponding to the change of Larmor precession frequency by as much as 56 MHz.

Figure 4(a) shows the schematic of the sequence used to demonstrate the phase rotation we propose. We incorporate ESR pulses to project the induced phase shift, since the PSB-based measurement is insensitive to the spin phase itself. First, the right spin (as well as the left spin) is initialized at a bias position M and then a step voltage is applied to fix an initial reference frame at $P_0$. In the following π/2-π-π/2 sequence, the three ESR pulses are equally spaced in time just as in a conventional spin echo sequence [11]. During the second interval a voltage pulse is applied to perform a phase gate at various bias points. The last X(π/2) rotation projects the spin phase to the *z*-axis.

The measured $\Delta G_{QPC}$ oscillates as a function of voltage pulse duration to accumulate the phase, as shown in Fig. 4(b). The Z gate frequency $f_Z$ ranges from 0 to 40 MHz from $P_0$ to $P_3$ (Fig. 4(c)), reflecting differences of local $B_z$. The maximum $f_Z$ of 54 MHz is obtained in a different condition (Fig. 4(d)). This $f_Z$ value corresponds to a 12 mT change of $B_z$, which can be accounted for by a shift of the electron of ~ 10 nm in the right QD. The time required for the commonly used π/8 gate, Z(π/4), is as short as 2.3 nsec. This is only half the gating time in the conventional sequence even with the 120 MHz ESR rotation obtained in this work. From numerical simulation the average gate fidelity [39] for Z(π) is estimated to be 98 %, assuming $T_2^Z = 36$ nsec.

We anticipate that within our scheme 200 MHz *x*- and *y*-rotations will be in experimental reach by using a thinner insulator (20 nm would be possible for instance with atomic-layer-deposition technology) to further reduce the distance between the MM and QDs. Undesirable PAT effects under strong MW burst may be suppressed by making the tunnel coupling more opaque during the ESR drive by gating or operating deeper in blockade with stronger QD confinement. We also expect that of the order of 100 MHz *z*-rotations will be feasible with optimized, larger pulses. Also, even higher operation fidelity may be obtained by improving readout fidelity with a rapid single-shot measurement technique [39]. The techniques demonstrated here should be readily applicable to other material systems with longer coherence times, e.g. group-IV semiconductors [24,40,41]. The large control fields (~ 20 mT) achieved here can implement single-qubit π rotations within 1 nsec in Si-based QDs (with *g* ~ 2), suggesting a fault-tolerant single-qubit gate fidelity [42,43] given the observed $T_2^* = 360$ nsec [41]. The exponential coherence decay observed here may be of significance for quantum error correction.

To summarize, by utilizing large inhomogeneous magnetic fields with a MM we have realized accurate spin flips up to 127 MHz and demonstrated a novel electrical phase control up to 54 MHz. These results will allow for high-fidelity single-qubit gates in large-scale quantum circuits under the premise that all operation-times are at least an order-of-magnitude shorter than the decoherence time. In the *fast* Rabi regime with $f_{\text{Rabi}} \gg \sigma$, we have observed distinct features of spin dynamics which indicate decoupling from the nuclear-spin bath, and revealed the difference of electron-nuclear spin dynamics between the *slow* and *fast* Rabi driving which has never been studied in bulk semiconductor ESR.

**Acknowledgement**

We thank G. Allison for carefully reading the manuscript. Part of this work is financially supported by a Grant-in-Aid for Scientific Research on Innovative Areas (No. 21102003) and Project for Developing Innovation Systems from the Ministry of Education, Culture, Sports, Science and Technology, Japan (MEXT), the Funding Program for World-Leading Innovative R&D on Science and Technology (FIRST), the IARPA project "Multi-Qubit Coherent Operations" through Copenhagen University, Grant-in-Aid for Scientific Research(S) (No. 26220710) from Japan Society for the Promotion of Science (JSPS), and ImPACT Program of Council for Science, Technology and Innovation (Cabinet Office, Government of Japan). M.P.-L. acknowledges financial support from the National Science and Engineering Research Council of Canada (NSERC) and the Canadian Institute for Advanced Research (CIFAR). J.Y. and T.T. acknowledges the financial support from JSPS for the Research Fellowships.

J.Y. and T.Ot. contributed equally to this work.

FIG. 1 (a) Scanning electron micrograph of a similar device along with the coordinate system. Shown in false-color orange is the shape of a 250 nm thick Co MM. High frequency pulses are applied to gates C and R, and MW solely to C. (b) Numerically simulated distribution of the stray field in the $x$-direction. The origin is at the center of the two QDs. (c) Numerically simulated distribution of the stray field in the $z$-direction.

FIG. 2 (a) CW-ESR signals as a function of $B_{ext}$ and $f_{MW}$ under CW MW irradiation. The peak at higher (lower) frequency comes from the right (left) spin, based on the MM field distribution. The g-factors $|g| = 0.333 \pm 0.006$ for both peaks. (b) An example of ESR spectra in the PP ESR, averaged over 50 $B_{ext}$ sweeps to minimize the effect of DNP. (c) Pump positions used for local Zeeman field probe in the stability diagram under ~ 500μeV source-drain bias. The purple line indicates the (1,1)-(2,0) charge boundary. (d) ESR peak field ($B_{ext}$) dependence on pump positions.

FIG. 3 (a) *Fast* Rabi oscillations measured at $B_{ext}$ = 0.51 T with $f_{MW}$ = 3.0 GHz. The estimated MW power at the sample is stepped by 3 dB from -32 to -11 dBm and in this range $f_{Rabi}$ ranges from 29 to 126 MHz. Solid lines are the fit to $C + A \exp\left(-(t_{MW}/T_2^{Rabi})^2\right)\cos(2\pi f_{Rabi} t_{MW})$ with $f_{Rabi}$ and $T_2^{Rabi}$ as fitting parameters. Traces are normalized so that $A = C = 0.5$ and offset for clarity. See the main text for the source of decay in this regime. (b) MW amplitude dependence of $f_{Rabi}$ extracted from (a), with a linear fit for data points of the four smallest powers with zero intercept. (c) FWHM of ESR spectrum v.s $f_{Rabi}$ under CW (red squares) or PP MW irradiation (red circles). Note the PP-ESR FWHM is for the maximum spin-flip signal, and can be smaller than the CW-ESR FWHM with the same $f_{\text{Rabi}}$ by at most 22%. The red line shows the fit to $(h/|g|\mu_B)\left[1.07 f_{\text{Rabi}} + \sqrt{0.858 f_{\text{Rabi}} + 4\ln 2 / (\pi T_2^*)^2}\right]$, which approximates the FWHM dependence of the Voigt profile [44]. Here $h$ is Planck's constant and $\mu_B$ is the Bohr magneton. The dotted black line indicates the $f_{\text{Rabi}}$ contribution. $f_{\text{Rabi}}$ of CW-ESR is estimated from the MW amplitude dependence in (b). (d) Comparison of different fit functions of 85.9 MHz Rabi data. The blue trace is the least-square fit by the *fast* Rabi expression, whereas the red one is by the *slow* Rabi approximation, $C' + A' t_{\text{MW}}^{-0.5} \cos(2\pi f'_{\text{Rabi}} t_{\text{MW}})$. Exponential decay gives the better fitting result with $f_{\text{Rabi}} = 85.9$ MHz and $T_2^{\text{Rabi}} = 36.7$ ns. (e) Same type of comparison as in (d) for the Rabi data with $f'_{\text{Rabi}} = 14.6$ MHz. Here only ESR peaks are collected and plotted to resolve slow Rabi oscillations [14,21]. (f) Intensity plot of the 85.9 MHz Rabi oscillation as a function of $B_{\text{ext}}$ and $t_{\text{MW}}$. A "chevron" interference pattern reflects vanishing influence from the nuclear-spin bath. Each data pixel is integrated for about 1 sec. (g) Same type of plot as in (f) for the 14.6 MHz Rabi data. The ESR peak broadening is predominated by the driving Rabi field of about 3 mT, rather than the Overhauser field fluctuation.

FIG. 4 (a) Schematic showing a gate pulse sequence for phase-gate demonstration, with expected spin orientations in the Bloch sphere representation with 62.5 MHz ESR rotations. (b) Oscillations demonstrating phase-gate operations for various gate pulse amplitudes indicated in (c), plotted in the same color code. Traces are fit to $C + A \exp\left(-(t_{\text{gate}}/T_2^{\text{ZGate}})^2\right) \cos(2\pi f_Z t_{\text{gate}})$. $\Delta G_{\text{QPC}}$ signal is linearly converted to probability from independent Rabi measurement. (c) A series of bias points used in (b) with different values of local $B_z$ at $P_0$ to $P_3$. (d) The fastest 54 MHz phase oscillation obtained in a different condition from that in (b).

Figure 1

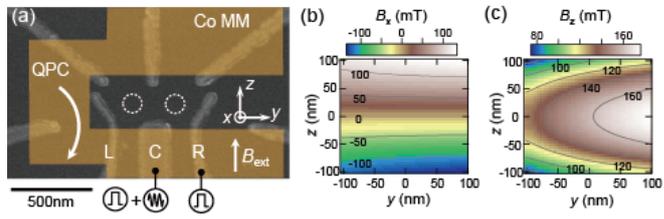

Figure 2

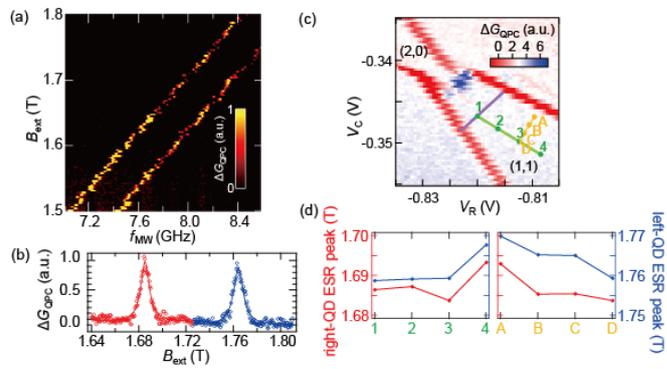

Figure3

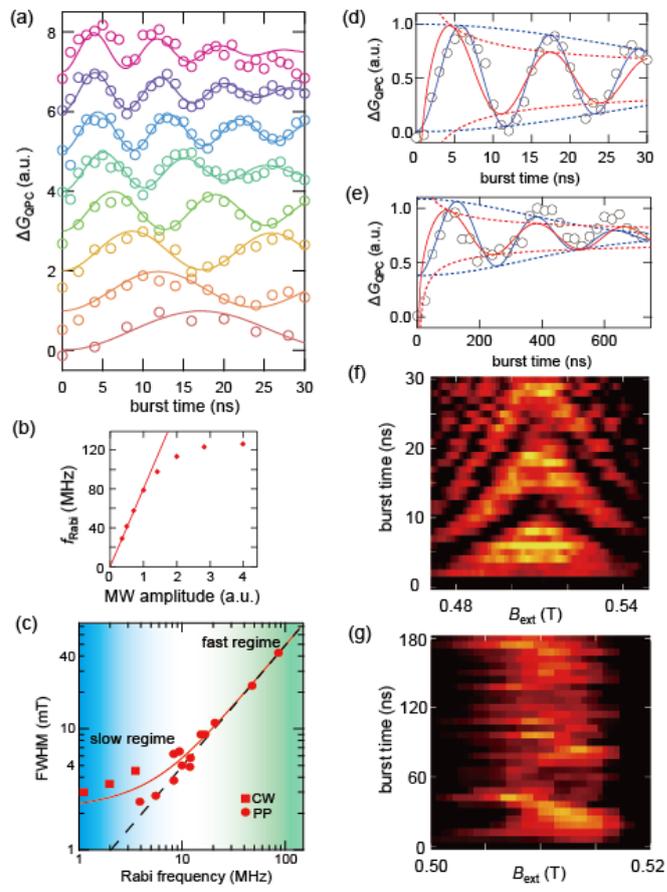

Figure 4

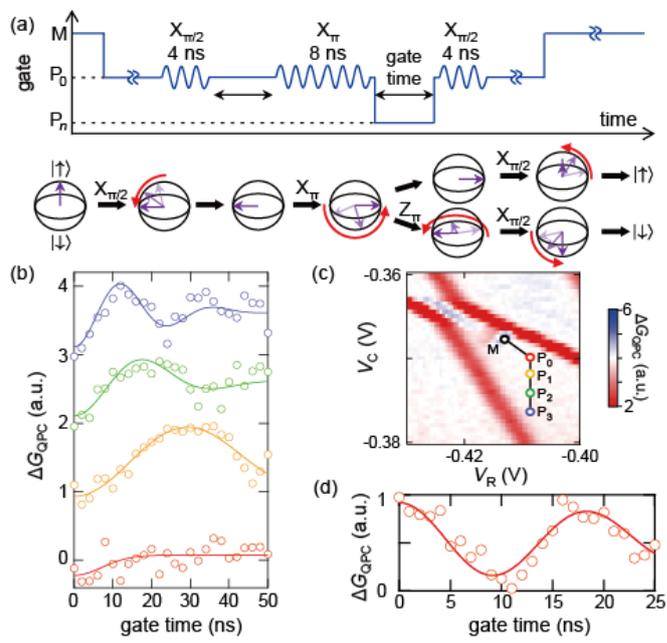

**Supplemental Material for Fast Electrical Control of Single Electron Spins in Quantum Dots with Vanishing Influence from Nuclear Spins**

1. Methods

The stray magnetic field produced by a cobalt MM is simulated using the Mathematica package Radia (available at http://www.esrf.fr). The 250 nm thick cobalt MM is deposited using a standard electron beam evaporator at a rate ~ 0.1 nm/sec. The surface gates are separated from the MM by a 100nm thick insulator (Calixarene). We choose the *z* axis along the crystallographic direction of [110] because the spin-orbit interaction then provides a local a.c. magnetic field to be added to $b_\text{sl}$ for the ESR rotation [28]. In addition, to make $E_\text{ac}$ as large as possible the two QDs are designed to be parallel-coupled to a gate electrode C to which MW is irradiated because PAT imposes an upper limit of $E_\text{ac}$ for the two QDs tandem-coupled to the MW gate [14]. $b_\text{sl}$ in this study is estimated to be roughly 3 times larger than in the previous report [15,17,21]. To account for the 10-fold enhancement of $f_\text{Rabi}$, another factor of 3 is attributed to 3-fold larger $E_\text{ac}$ as a result of the parallel-coupled configuration. A trans-conductance technique was used throughout, where the 216 Hz excitation is applied to the gate R.

2. Chevron patterns

When the Rabi frequency ($f_\text{Rabi} = \omega_\text{R}/2\pi$) and the MW frequency detuning ($\Delta\omega/2\pi$) are both much smaller than the resonance frequency, the rotating wave approximation holds and the ESR spin flip probability $P(t_\text{MW}, \Delta\omega)$ can be expressed using the Rabi formula:

$$P(t_\text{MW}, \Delta\omega) = \frac{\omega_\text{R}^2}{2[\omega_\text{R}^2+\Delta\omega^2]}\left[1-\cos\left(\sqrt{\omega_\text{R}^2+\Delta\omega^2}\,t_\text{MW}\right)\right], \qquad (1)$$

where $t_\text{MW}$ is the MW burst duration. A map of $P(t_\text{MW}, \Delta\omega)$ makes a chevron pattern in the time-spectral ($t_\text{MW}$-$\Delta\omega$) domain. When we take into account the effect of static Gaussian-distributed Overhauser field, the pattern becomes blurred;

$$\overline{P}(t_\text{MW}, \Delta\omega) = \int_{-\infty}^{\infty} d\delta \frac{1}{\sqrt{2\pi}\sigma_\omega} e^{-\frac{\delta^2}{2\sigma_\omega^2}}\, P(t_\text{MW}, \Delta\omega + \delta).$$

Supplementary Figure 1 plots maps of calculated $\overline{P}(t_\text{MW}, \Delta\omega)$ for different Rabi frequencies. The chevron pattern can be recognized only when $\omega_\text{R}/\sigma_\omega \gtrsim 4$ or $f_\text{Rabi} \gtrsim 28$ MHz with a typical $\sigma_\omega/2\pi \equiv \sigma = 7.0$ MHz for GaAs QDs [11].

3. Different fit functions for *slow* and *fast* Rabi oscillations

When the time averaged detuning is zero, as is intended in usual experiments, the average spin flip probability reads

$$\overline{P}(t_{MW}) = \int_{-\infty}^{\infty} d\delta \frac{1}{\sqrt{2\pi}\sigma_\omega} e^{-\frac{\delta^2}{2\sigma_\omega^2}} \frac{\omega_R^2}{2(\omega_R^2+\delta^2)} \left[1 - \cos\left(\sqrt{\omega_R^2 + \delta^2}\, t_{MW}\right)\right]. \quad (2)$$

In the weak spin-flip drive with $\omega_R \ll \sigma_\omega$, for $t_{MW} \gg 1/\omega_R$,

$$\overline{P}(t_{MW}) \sim C - \frac{1}{2}\sqrt{\frac{\tau}{t_{MW}}} \cos\left(\omega_R t_{MW} + \frac{\pi}{4}\right), \text{ with } \tau = \frac{\omega_R}{2\sigma_\omega^2}$$

whereas in the strong spin-flip drive with $\omega_R \gg \sigma_\omega$,

$$\overline{P}(t_{MW}) \sim \frac{1}{2} e^{-(t_{MW}/2\tau)^2} [1 - \cos(\omega_R t_{MW})],$$

as expected for pure dephasing. From Supplementary Figure 2 we find that the Rabi oscillation suffers from an initial phase shift and reduced amplitude for $\omega_R/\sigma_\omega < 2$ in the *slow* Rabi regime.

4. Rabi fidelity estimation

In the main text we estimated ESR spin flip fidelity for up/down spin input states from Rabi oscillations. The ESR signal is scaled to 0 at the initial value of measured transconductance signals, $\Delta G_{QPC} = dG_{QPC}/dV_R$ and 1/2 at its setting value. This procedure is valid in the strong spin flip drive, because for $\omega_R \gg \sigma_\omega$, the spin flip probability tends to 1/2. Indeed, from Eq. (1), one can calculate the final spin flip probability as

$$\overline{P}(t_{MW} = \infty) = \frac{\sqrt{\pi}\omega_R}{2\sqrt{2}\sigma_\omega} e^{\frac{\omega_R^2}{2\sigma_\omega^2}} \left(1 - \text{Erf}\left(\frac{\omega_R}{\sqrt{2}\sigma_\omega}\right)\right) \approx \frac{1}{2}\left(1 - \left(\frac{\omega_R}{\sigma_\omega}\right)^2\right). \quad (3)$$

Therefore, $\overline{P}(t_{MW} = \infty) \sim 0.5$ for $\omega_R/\sigma_\omega > 5$.

**Supplementary Figure 1** | Spin flip probability or Rabi oscillation as a function of normalized detuning and burst time calculated for different Rabi frequencies with the ratio of $\omega_R/\sigma_\omega = 1$ (a), 2 (b), 4 (c), 8 (d), and $\infty$ (e), respectively.

**Supplementary Figure 2** | Numerical calculation of spin flip probability or Rabi oscillation for the ratio $\omega_R/\sigma_\omega$ ranging from 0.5 to 10 in steps of 0.5 using Eq. (2). The burst time is normalized by the Rabi period. Dotted lines are guides to the eye for the initial $\pi/4$ shift in the Rabi oscillation.

Supplementary Figure 1

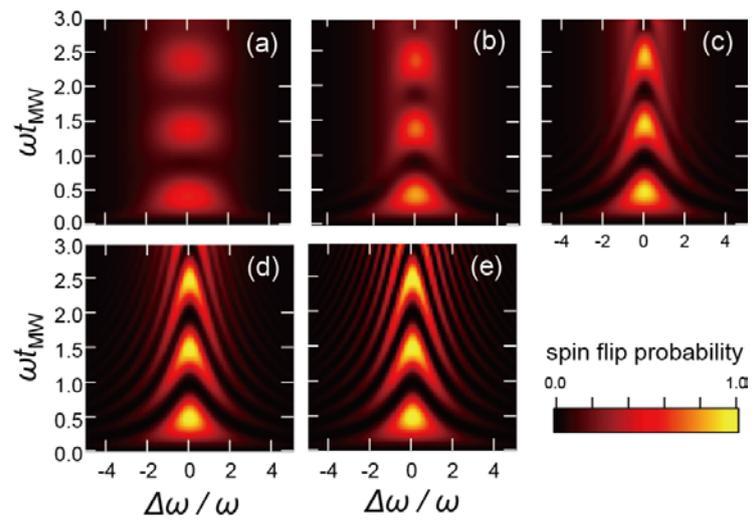

Supplementary Figure 2

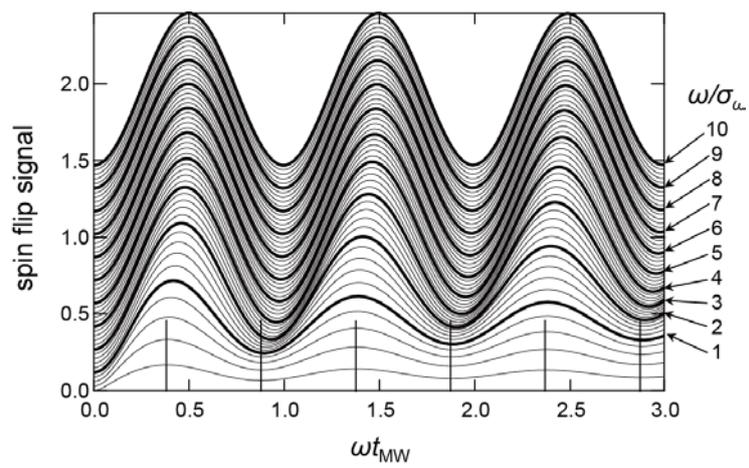